\documentclass[review]{elsarticle}

\journal{Journal of \LaTeX\ Templates}

\begin{document}

\begin{frontmatter}

\title{The Gaia Data Release 1 parallaxes and the distance scale of Galactic planetary nebulae.}

\author[a]{Letizia Stanghellini}

\address[a]{National Optical Astronomy Observatory,
   950 N. Cherry Avenue,
            Tucson, Arizona 85719 (USA)}
\author[b,c]{Beatrice Bucciarelli}
\author[b]{Mario G. Lattanzi}
\author[b]{Roberto Morbidelli}

\address[b]{INAF, Osservatorio Astronomico di Torino, Via Osservatorio, 20, 10025 Pino Torinese TO, Italy}

\address[c]{Shanghai Astronomical Observatory, Chinese Academy of Sciences, 80 Nandan
Rd, 200030 Shanghai, China}

\begin{abstract}
In this paper we gauge the potentiality of Gaia in the distance scale calibration of planetary nebulae (PNe) by assessing the impact of DR1 parallaxes of central stars of Galactic PNe (CSPNe) against known physical relations. 
For selected PNe targets with state-of-the-art data on angular sizes and fluxes, we derive the distance-dependent parameters of the classical distance scales, i.e., physical radii and ionized masses, from
DR1 parallaxes; we propagate the uncertainties in the estimated quantities and evaluate their statistical properties in the presence of large relative parallax errors; we populate the statistical distance scale diagrams with this sample and discuss its significance in light of existing data and current calibrations.
 We glean from DR1 parallaxes 8 CSPNe with S/N$>$1.
We show that this set of potential calibrators doubles the number of extant trigonometric parallaxes (from HST and ground-based), and increases by two orders of magnitude the domain of physical parameters probed previously. We then use the combined sample of suitable trigonometric parallaxes to fit the physical-radius-to-surface-brightness
relation. This distance scale calibration, although preliminary, appears solid on statistical grounds, and suggestive of new PNe physics.
 With the tenfold improvement in PNe number statistics and astrometric accuracy expected from future Gaia releases the new distance scale, already very intriguing, will be definitively constrained.

\end{abstract}


\end{frontmatter}


\section{Introduction}

Planetary nebulae (PNe) are the dust and gas shells ejected at the late AGB stellar phases, and then ionized, partially or fully, by the radiation from the hot, evolving central stars (CSs). PNe are probes of stellar and nebular evolution and of cosmic chemical enrichment, and knowing their formation and evolution is essential in many astrophysical fields.

To date, several hundred Galactic PNe are known (e.~g., catalogs by Acker et al.~1992; Parker et al.~2006), but their distances have always been elusive. Reliable independent distances are known for a relatively small number of Galactic PNe (Stanghellini et al.~2008, hereafter SSV) if we exclude those whose distances are model dependent (Frew et al.~2016, hereafter F16). Since the PN distance is needed to study the astrophysics of the nebulae and their CSs, scientists in the field have recurred to calibrate physical relations between distance-dependent and distance-independent astronomical parameters of the PNe. Once these relations have been calibrated for the few PNe with a credible independent distance determination, they are used as a distance scale, yielding distances to all Galactic PNe whose distance-independent parameters used in the scale can be measured.

The Galactic PN distance scales are typically derived by using a measure of the PN surface brightness (or its inverse, the optical thickness) as the independent parameter, and with the physical radius (e.~g., Schneider \& Buckley 1996; Shaw et al.~2001; F16) or the ionized mass (e.~g., Daub 1982; Cahn et al.~1992, SSV) as distance-dependent parameter. The original idea of the statistical distance scale is due to Shklovskii (1957). It is beyond the scope of this paper to review all the literature regarding the development of distance scales, but there is purpose in reviewing the basic concepts. 

It is assumed that a set of several PNe whose physical parameters, including distance, have been accurately measured, offers a snapshot of PN evolution. In the first method (which we will refer to as {\it physical radius} distance scale), it is assumed that PN surface brightness decreases with time since PN ejection, while the physical radius increases; in the second scenario (the {\it ionized mass} distance scale), the ionized mass increases for optically-thick PNe, while it stays approximately constant for optically-thin PNe, as the PN evolves. Both scales have their drawbacks and advantages (see a detailed discussion of the scales in SSV and F16), but they are meaningless unless their calibrators are spot-on reliable. 

Good PN distance calibrators are rare. Stanghellini et al.~(2008, their Table 2) list all independent distances of PNe that are model independent and reliable, although the distance uncertainties may be large, and in some cases they are not available at all from the original distance measurements, which makes the assessment of their quality hard. Of the many calibrators that have been used to date, the most reliable are those whose distances can be measured from trigonometric parallaxes, cluster membership of the PNe, and from spectroscopic parallaxes, usually determined by using stellar properties such as the presence of a companion to the CS. Secondarily, one can employ distances derived from the expansion of the nebulae; this method is not as accurate as the previous ones, since it  does not account for the possible acceleration of the nebular ejecta. PN distances can be estimated also from extinction of the PN itself, and of a selection of nearby stars, by building an extinction-to-distance relation for each PN; the extinction method is biased by patches in the ISM extinction, which are hard to predict, and by a mild model dependency of the stellar distance scale. 

Given the limited number of reliable calibrations, SSV recurred to Magellanic Cloud PNe as calibrators. There are many dozen of such PNe  observed with the {\it HST}, which allows to measure their apparent radii with very low uncertainty (Magellanic Cloud PNe are typically unresolved from the ground). The SSV calibration is probably the best scale for most PN distances; in fact, it is the one that best reproduces the independent distances from spectroscopic parallaxes and cluster membership observations. 
 A possible source of error for this scale is that Magellanic Cloud PNe become optically thin at higher surface brightness than Galactic PNe, due to the lower metallicity of the former compared to the latter. Thus by using Magellanic Cloud PNe as distance scale calibrators one may overestimate distances of optically thick Galactic PNe.

The situation of Galactic PN distances is not going to improve considerably unless we increase the number of calibrators, their quality, or if we could reduce the calibrator's distance error bars. The first Gaia Data Release (DR1; Gaia Collaboration, Brown A., Vallenari A. et al.~2016) offers the opportunity to test new directions for Galactic PN distance scales. Gaia measured the parallaxes of several CSs of nearby PNe, and we can use the few good parallax measurements of the DR1 as an initial tool to explore the PN distance scales with the {\it physical radius} and {\it ionized mass} methods.

In $\S$2 we describe the PNe in the DR1 sample. In $\S$ 3 we calculate the calibration for the  {\it physical radius} and {\it ionized mass} distance scales using DR1 parallaxes and other trigonometric parallaxes in the literature as calibrators. In $\S$4 we discuss the results. Finally, $\S$5 gives the conclusions of our study, and its foreseen future developments.

\section{A sample of Galactic PNe with a CS parallax in DR1}

\subsection{Searching DR1 for CSPNe}

Trigonometric parallaxes for a set of about 2 million stars, mostly brighter than 11.5 visual magnitude, are delivered by DR1 (the so called {\it primary data set},  Lindegren et al.~2016, or {\it TGAS solution}, Michalik et al.~2015).  This task has been accomplished by combining 14 months of Gaia observations with earlier positions from the Hipparcos and Tycho-2 catalogs, thereby allowing to
 disentangle  the component of translational motion from the parallactic one, at the same time preserving the independent and absolute nature of parallax estimations.  As detailed in Lindegren et al.~(2016), the typical parallax uncertainty of DR1 stars is $\approx 0.3$ mas; only  sources having a formal parallax error smaller than 1 mas were retained in this release. At the end of the nominal five-year mission, however, Gaia is expected to deliver an order of magnitude or more improvement for these sources, while providing parallaxes with sub-mas precision even for objects as faint as V$\sim$20.
 
We searched the DR1 dataset for stellar detection and parallax determination corresponding to the locations of CSs of Galactic PNe. We used as input  the astrometry in Kerber et al.~(2003) of all spectroscopically confirmed PNe in the Galaxy, as listed in the Strasburg-ESO catalog (Acker et al.~1992), combined with the MASH survey (Parker et al 2006).  

In Table 1 we list all detected PNe whose CS TGAS parallax measurement was available. We give the PN~G number (column 1), the common name of the PN (2), the Gaia ID of the CS observation (3), the CS parallax, $p$ and its uncertainty $\sigma_{\rm p}$ in mas,  from DR1 (4), and the logarithmic distance and its uncertainty, in parsecs (5).  Logarithmic distances estimates are directly obtained from DR1 parallaxes as ${\rm log}(D_{\rm p})=3-{\rm log}(p)$ (since the parallax is in mas and the distance in parsec), and the corresponding
asymmetric errors bars computed from the formal parallax variances as ${\rm log}[(p+\sigma_p)/p]$, ${\rm log}[p/(p-\sigma_p)]$ for the lower and upper limits,
respectively.

Given the paucity of PNe targets in DR1, we decided to include in this explorative analysis all of the objects with S/N $>$ 1, which is the threshold for the presence of signal. Large relative parallax errors must be handled carefully and this will be addressed whenever relevant in the following sections.

In Table 1 we do not include three targets that were found in this search but have been deemed to be misclassified PNe by F16: PN~G050.1+03.3 is a WR ejecta, PN~G288.9-00.8 a LBV ejecta,  and PN~G303.6+40.0 is a patch of ionized ISM.  It is worth noting that PN~G334.8-07.4 (SaSt2-12) is a halo PN (P04), and possibly a post-AGB star rather than an evolved PN. We include it in the sample, and flag it, when relevant, in the analysis below. 

\subsection{Building the sample of PN calibrators}

We matched the Gaia CS detection in DR1 against the PN images to infer whether the Gaia detections correspond to the PN CSs. In order to do so we use optical PN images in the literature, giving preference to those in the MAST archives, not only because  {\it HST} images have the best resolution in the optical wavelengths to date, but especially since their astrometry is compatible with that of Gaia, and the comparison can be quantitative.  In a complementary paper we will describe these techniques, and show the Gaia to {\it HST} correspondence for all PNe released by Gaia in DR1 and imaged by {\it HST}, including those whose parallax is not measured at this time.  For all PNe used here, the CS location in the optical images corresponds to the Gaia position for the parallaxes. 

In order to determine the distance scale parameters we need the $H\beta$  fluxes, $F_{\rm H\beta}$, their extinction corrections, the apparent angular radii, and the 5 GHz fluxes for all calibrators. In Table 2 we give, for each target with a determined TGAS parallax distance, the usual PN name (column 1), the optical angular radius in arcsec (2) the $H{\beta}$ flux in erg cm$^{-2}$ s$^{-1}$ (3),  the logarithmic optical extinction constant (4), and the 5 GHz flux in Jy (5). In some cases, where the 5 GHz fluxes were not available, we used the transformation by Cahn et al.~(1992) to determine them  from the $H\beta$ fluxes, as noted in Table 2.

Uncertainties for all parameters are given when available in the original references. When $H\beta$ flux or extinction uncertainties were not available in the references, we assumed them to be 0.02 in the log, while unavailable uncertainties in the 5 GHz flux were assumed to be 
10$\%$ of the flux; these are typical uncertainties for the corresponding parameters (Cahn et al.~1992). We searched the literature for new measurements of fluxes and angular diameters of the nebulae since the work by Stanghellini \& Haywood (2010, hereafter H10).  Pereira et al.~(2010) measured the angular dimension of the inner, highly emitting parts of PC~11, based on HST images. We use their determination, assuming an error bar which is reasonable given the asymmetry of the PN.  For SuWt~2, we found a measure of the $H\alpha/H\beta$ flux ratio (Danehkar et al.~2013), which, together with the $H\alpha$ flux measured by Frew et al.~(2013), gives log($F_{\rm H\beta}$)= -12.35, and extinction constant c=0.64. We use this flux to obtain an equivalent 5 GHz flux of 0.007 Jy, to be used in the distance scale formulation. All other parameters in Table 2 have been taken from SH10.
  
Fluxes and radii are available from the
literature for all 8 PNe whose parallax distance is available in DR1 with S/N $>$1; however given this low number of targets and the usually low relative parallax
accuracy, it is obvious that this sample can only be adequate to build a preliminary distance scale. Moreover, the nature of PNe as astrophysical objects, and the fact
that they are often associated with varying sky background, make them
potentially difficult targets for astrometry. Therefore, by analyzing the
location of Gaia DR1 parallaxes on the classical PNe statistical distance
scales we can in turn test their astrometric quality, and at the same time draw
some conclusions on the prospects of future developments with the foreseen Gaia releases.

  \begin{figure}
   \centering
  \includegraphics[width=\hsize]{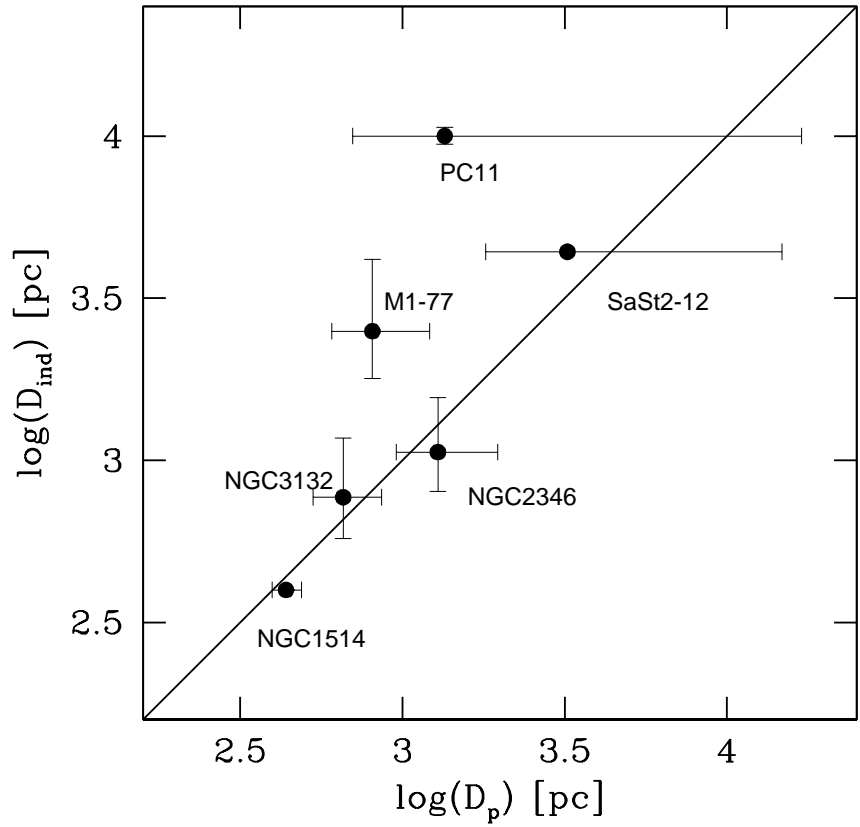}
\caption{Comparison between independent distances of Table 3 vs. trigonometric distances from DR1. The solid line is the 1:1 relation.}
    \end{figure}

\subsection{Gaia parallaxes of Galactic PNe compared to other independent distances}

Before we use the TGAS parallaxes, let us compare their derived distances with other distance measurements that are in the literature to date. In Table 3 we list the available independent distances, and their uncertainties, together with the determination method and references, for those PNe with TGAS parallaxes whose independent distance was available.  In principle, we could select independent distances derived from trigonometric and spectroscopic parallaxes, group membership, nebular expansion, and extinction distance methods. 

The trigonometric parallax distances are the most reliable ones, but none of the nearest PNe with known trigonometric parallaxes (Smith 2015) were found in the DR1 list, as all these objects have magnitudes fainter than those in DR1. Most of the comparison distances in Table 3 are thus from spectroscopic parallaxes, the second best method. Reddening (or extinction) distances are the next choice when a parallax is not available, such as for M~1-77.  For NGC~2346, Mendez \& Niemela (1981) have estimated the distance from the spectroscopic binary located at the PN center, but incompatible extinction determinations of the CS and the nebula make it unclear whether the stellar distance is really the nebular distance. For this reason, we do not use in the following the spectroscopic parallax for NGC~2346, rather, we use its reddening distance (see Table 3).

We did not include in this comparison the strongly model-dependent distances, such as those that assume a given CS mass (e.~g., Herald \& Bianchi 2011), gravity (e.~g., Maciel \& Cazetta 1997), SED  (e.~g., Vickers et al.~2015), and all other model-dependent distances.  It is worth noting that there is a mild model dependency of spectroscopic distances as well, since the binary (or multiple) CS is modeled to infer the stellar type of the bright stellar companion to the ionizing source. Nonetheless, they are still the most reliable distances besides trigonometric parallaxes.

The PN NGC~1514 has a spectroscopic parallax determination from Aller et al. (2015), giving ${\rm log}(D_{\rm pc})=2.403^{\rm +0.130}_{\rm -0.186}$, but in Table 3 we use instead the distance inferred by Mendez et al. (2016) that is needed to reconcile the equal distance of the two binary central star components, which is physically more interesting for this system.

As a sanity check, in Fig. 1 we plot the independent distances of Table 3 against distances from TGAS parallaxes, as in Table 1. Asymmetric uncertainties have been plotted when available. Uncertainties associated to independent distances are from the original references, given in Table 3.  Four targets are within the stated 1$\sigma$ error, and two are compatible with the 1:1 relation within 2$\sigma$. 
If we were to perform a least square fit based on these data, the
parallax logarithmic distances would set $\sim$0.1 dex below the independent ones. Given the limited number of objects and the relatively high errors involved, we do not attempt to explain this discrepancy. Nonetheless, Fig. 1 offers a  compelling comparison between PN distances from Gaia parallaxes and other methods.

\section{Galactic PN distance scale based on Gaia parallaxes}

\subsection{The physical radius-optical thickness distance scale}
\begin{figure}
   \centering
  \includegraphics[width=\hsize]{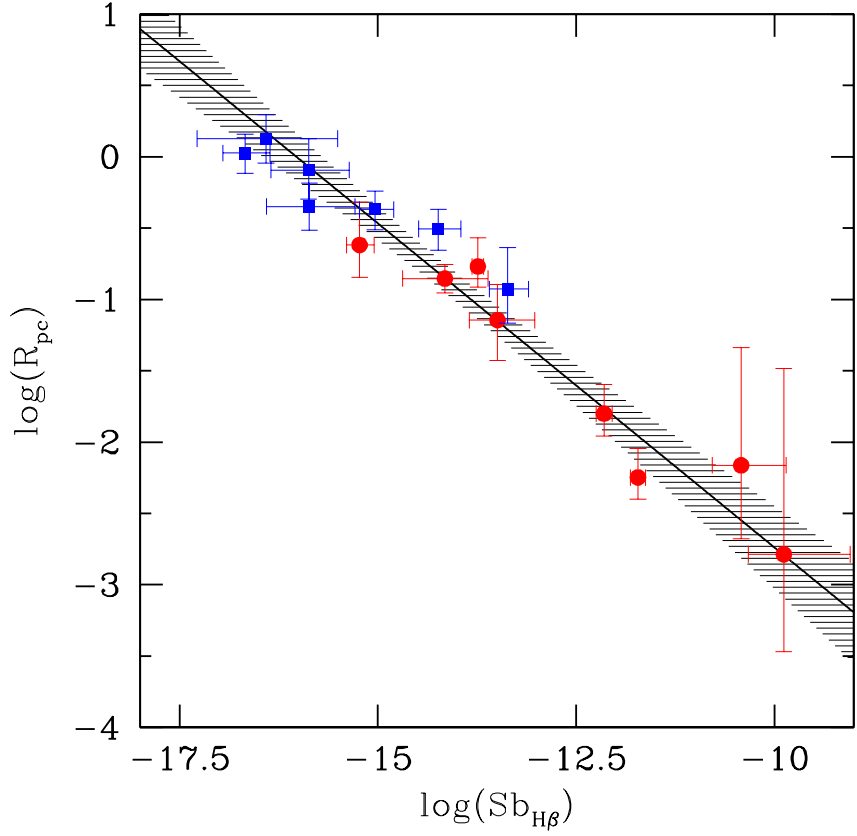}
\caption{The physical radius-surface brightness relation. Filled red circles: PNe with Gaia DR1 parallaxes. Filled blue squares: PNe with trigonometric parallaxes from the literature. Error bars represent the 1$\sigma$ asymmetric uncertainties, as explained in the text. The solid line represents the linear fit to all trigonometric parallax data, taking into account averaged uncertainties, and excluding SaSt~2-12. The shaded area represent the 1$\sigma$ confidence region of the fit (see text).}
\end{figure}

In Figure 2 we plot the PNe with TGAS parallax distances (filled red circles) in the physical radius vs. surface brightness logarithmic plane. ${\rm log}(R_{\rm pc})={\rm log}(\theta/(p\times206.265))$ is derived from the TGAS parallaxes and the optical angular radii. The $H\beta$ surface brightness is ${\rm log}(Sb_{\rm H\beta})= {\rm log} (I_{\rm H\beta}/\pi \theta^2$), where ${\rm log}I_{\rm H\beta}={\rm log}F_{\rm H\beta}+c$ is the extinction-corrected logarithmic $H{\beta}$ flux. The data points show a good linear correlation, despite some large error bars coming especially from the high uncertainties in DR1 parallaxes. We attempt to estimate the parameters of the expected linear ${\rm log}(R_{\rm pc})-{\rm log}(Sb_{\rm H\beta})$ relation, with particular caution on the statistical handling of the errors.

In order to study the influence of TGAS errors on the estimation of the PN physical radius, the physical quantity
underlying the distance scale relation,  we tested the non-linearity and asymmetry of our estimator ${\rm log}(R_{\rm pc})$ by means of Monte Carlo-like simulations.
By sampling the probability distribution of  ${\rm log}(R_{\rm pc})$ as function of the observed parallax $p$, we took the quantity
$10^{<{\rm log}(R_{\rm pc})>}/(R_{\rm pc, true})$ 
as empirical estimate of the bias introduced by the logarithmic transformation in the presence of relative parallax errors as large as 90$\%$, obtaining ratios
larger than 1, with a maximum deviation of $\sim$13$\%$, that we consider acceptable for the present analysis.

As for the propagated uncertainties, we computed the 68$\%$ (1$\sigma$)
intervals of the ${\rm log}(R_{\rm pc}$)  estimate and compared them with its formal standard deviation evaluated up to the second-order Taylor expansion. The ratios of the left and right asymmetric errors with
the formal sigma is 0.6 and 2.1 respectively for $\sigma_{\rm p}=90\%$, decreasing to 0.9 and 1.1 for $\sigma_{\rm p}=30\%$. The error bars shown in the figures correspond to the 1$\sigma$ uncertainties in the parallaxes and the other measured parameters.

In order to extend the sample of calibrators, and to compare the domain spanned by TGAS parallaxes with that covered
by other trigonometric parallaxes, we plot in Figure 2 those PNe whose CS trigonometric parallax has been reliably measured in the literature (filled blue squares). We selected all PNe with reliable trigonometric parallaxes from the summary paper by Harris et al.~(2007; see also Smith 2015). We  excluded from our plots and analysis those targets whose nature as PN has been dismissed (DeHt~5,  Re~1, TK~2,  PHL~932, F16; HDW~4, de Marco et al.~2013); we also excluded PG~1034+001, since observations to determine its parallax have been deemed insufficient (Harris et al.~2007). Angular radii, $H\beta$, extinction, and 5 GHz fluxes of PNe hosting the CSs with parallaxes in Harris et al.~(2007) have been taken from Cahn et al.~(1992), Acker et al.~(1992), Pottasch et al.~(1996), and Zhang (1995). Uncertainties in the angular radii were not available for these PNe in the literature; we have assumed 20$\%$ errors in the angular measurements, which is conservative especially for the large PNe of this sample. Flux uncertainties were usually available, otherwise, we made the same assumptions we did for the Gaia parallax sample (see $\S$2.2). In Table 4 we give the PNG number and the PN name (Columns 1 and 2), the parallax from Harris et al.~(2007) (3), and the other PN parameters used in the scale calibrations (4 --7). 
 
By comparing the domain probed by TGAS and the other calibrators in Figure 2 we note that TGAS parallaxes open up the domain of high surface brightness PNe, and extend the physical radius domain by a factor of $\sim$2 in the logarithm with respect to the earlier calibrators, potentially anchoring the relation for bright and compact PNe.

We fit these data points, and their uncertainties, by using the {\it fitexy} routine, available in {\it Numerical Recipes} by Press et al.~(1992, p. 1007). This routine fits a straight-line model to (x, y) data with (symmetric) errors in both coordinates. We used different combinations of the left and right error bars to test the sensitivity of the solution to different weights, obtaining a
maximum variation of the values of slope and intercept of 3$\%$ and 2$\%$
respectively. The handling of asymmetric errors in the fitting process is possible in principle, though numerical codes are not readily available, but its effectiveness must be investigated and we plan to do so when a
re-calibration of the PNe distance scale will become possible with the next Gaia data releases. By using the average of left and right error bars as weights we obtain from the fit:
$${\rm log}(R_{\rm pc})=-(0.454\pm0.043)\times{\rm log}(Sb_{\rm H\beta})-(7.274\pm0.610) \eqno(1)$$
plotted in Fig.~2 as a solid line,  which has $\chi^2$ probability q$\sim$0.4, and $\chi^2\sim$12, which is quite good compared to the 12 degrees of freedom for the fit (Press et al.~1992). Note that we excluded SaSt~2-12 from this fit  (see $\S$2.1).

\begin{figure}
   \centering
  \includegraphics[width=\hsize]{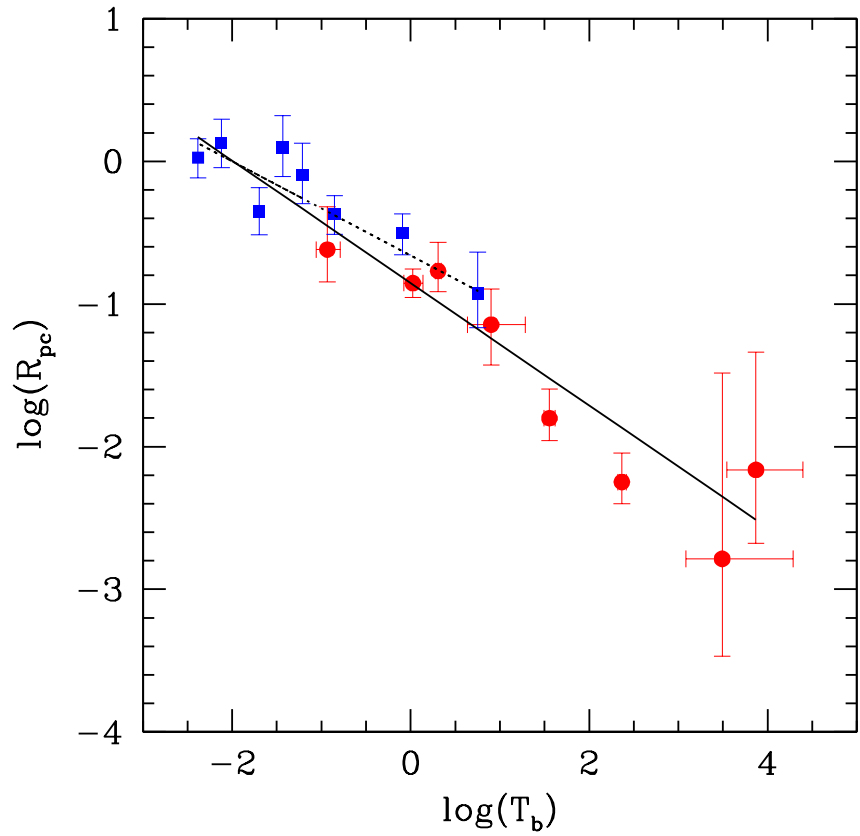}
\caption{The physical radius-brightness temperature relation. Symbols are as in Fig. 2. The solid line represents the linear fit to all trigonometric parallax data points, taking into account the uncertainties (see text), and excluding SaSt~2-12. The dotted line is the fit by Smith (2015), based on the Harris et al.~(2007) sample.}
\end{figure}

The correlation index between abscissae and ordinates of Fig. 2 is $R_{\rm xy}$ = -0.97, indicating almost perfect linear anti-correlation. In order to obtain che covariance of our slope and offset estimates, we used the {\it Gaussfit} software (Jefferys et al.~1988), which provided almost identical
estimates of the fit parameters of Eq. (1), and a correlation of $\sim99\%$. Using Monte Carlo methods, we then probed the $68\%$ error
ellipse of the parameter space to determine the confidence region of the fit, which is highlighted in Fig. 2.

While Eq. (1) can be used to determine distances to all Galactic PNe whose surface brightness is measured, it is too early to use it as a distance scale. Future Gaia releases will help setting this scale, which will probably be the most useful of all scales for PN distances in the future. 

Caution must be paid to the fact that we did not try to incorporate systematic errors as such in our analysis, mainly due to the small number
      statistics at play. However, as reported in Lindegren et al.~(2016), DR1 parallaxes are likely to be affected by systematic errors up to 0.3 mas (depending, e.g.,
      on star position and color), for this reason the final uncertainties in DR1 were artificially inflated. Besides, the size of such systematics could be in some cases significantly smaller than the formal uncertainties, as suggested by Casertano et al.
      (2017). In any case, the expected higher precision of future Gaia releases must be paralleled by the improvement in accuracy or else systematic errors 
     would seriously affect the results of any accurate statistical distance scale calibration.

In order to compare our data set with other distance scales in the literature we show, in Figure 3, both the TGAS (filled red circles) and other trigonometric parallax (filled blue squares) data sets in the physical radius vs. brightness temperature scale, where the brightness temperature is defined as $T_{\rm b}=17.5\times F_{\rm 5 GHz}/\theta^2$ (van de Steene \& Zijlstra 1995).  The brightness temperature is another way to express the optical thickness, but at radio wavelengths, similarly to the $\tau$ parameter that will be introduced in the ionized-mass scale in the next section. 
{All asymmetric uncertainties and fits have been dealt with as above in the physical radius-surface brightness calibration.
The resulting fit, excluding SaSt~2-12 (which has the highest T$_{\rm b}$), is 
$${\rm log}(R_{\rm pc})=-(0.429\pm0.041)\times{\rm log}(T_{\rm b})-(0.853\pm0.051),\eqno(2)$$  
with $R_{\rm xy}$=-0.96. It is worth comparing this fit with the equivalent scale of Smith (2015, their Fig. 9), which is also shown in Fig. 3. While the two fits broadly correspond in the overlapping domain, it is clear that smaller Gaia parallax uncertainties in the future will be essential to refine this PN distance scale.

\subsection{The ionized mass-optical thickness distance scale}

\begin{figure}
   \centering
  \includegraphics[width=\hsize]{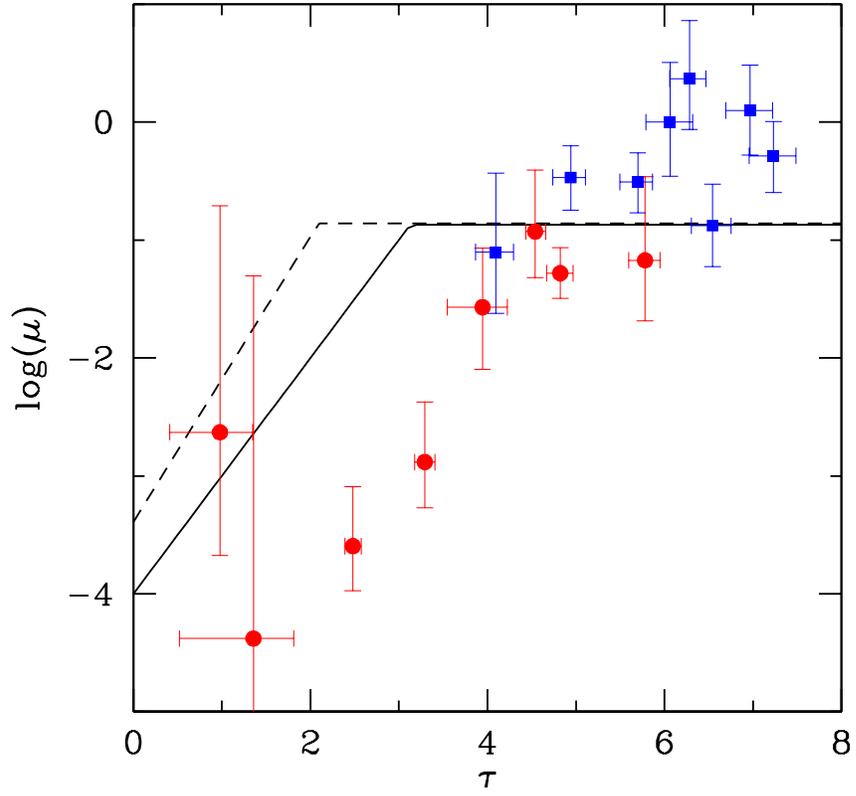}
\caption{The ionized mass--inverse  surface brightness relation.  Symbols are as in Fig. 2. The solid line is the Cahn et al.'s (1992) distance scale, while the broken line is the SSV scale calibrated on Magellanic Cloud PNe.}
\end{figure}

We show in Figure 4 the {\it ionized mass} vs. optical thickness distance scale. In the figure we plot
${\rm log}(\mu)={\rm log}(\sqrt(2.266~10^{-21} D^5 \theta^3 F_{\rm 5 GHz})$ (where $D$ is the PN distance in pc, $F_{\rm 5 GHz}$ is the flux in Jy, and $\theta$ is the angular radius in arcsec) versus $\tau = {\rm log}(4 \theta^2/F_{\rm 5 GHz})$.
This is the same formalism used by Cahn et al.~(1992), and SSV. The independent variable is a traditional distance scale variable, defined proportionally to the inverse 5 GHz surface brightness, while the dependent variable is the ionized mass of the PN. Filled red circles represent the set with TGAS parallaxes, while the squares are the other trigonometric parallax set. We plot only PNe whose $H\beta$ fluxes, their extinction constant, and their angular radii are available in the tables. 
Error bars are treated as above, showing the asymmetric bars for the 1$\sigma$ level. All parameter uncertainties have been included in the bars. Note that we do not fit these data points.
The solid (Cahn et al.~1992) and broken (SSV) lines are the state-of-the-art of the traditional distance scales, with their two-slope line tracing the optically-thick and optically-thin phases of PN evolution. During the optically-thick phase, the ionized mass increases as the nebula expands until $\tau$ reaches a $\tau_{\rm crit}$ and the nebula becomes completely ionized, at which point its ionized mass is assumed to remain constant.  It is worth noting that the lower extension of the traditional scales of SSV and Cahn et al.~(1992) is ${\rm log}(\mu)=-2$, indicating that the TGAS parallax set reaches to lower ionized masses, corresponding to early evolution PNe.  Since DR1 provides parallaxes for bright PNe only, we expect that the wealth of parallaxes coming from successive Gaia releases will allow for the first time to explore the optically-thick sequence in such detail that could revolutionize the calibration of this distance scale.

It is intriguing that a few of the optically-thick PNe with CS parallaxes from Gaia, whose ionized masses are spread over a range of values, seem to define a sequence that is different, albeit
parallel, to those of the older scales. The upper limit of the ionized mass  of the new scale is broadly consistent with that of the old scales. 

The possible new optically-thick sequence of Figure 4 is very interesting, insofar as it implies that the final ionization is reached at a later time (i.e., higher $\tau$) than indicated by the old scales, supporting the perception that there is more than one optically-thick evolutionary track in this plane as suggested by SSV (see Fig.  3 therein). This may signify that there is a delay in the thinning of the ejecta. It is worth noting that the only PN of the TGAS parallax sample that does not seem to follow this possibly new relation is SaSt~2-12 (the filled circle with the lowest $\tau$ in the Figure) which seems to be more in agreement with the Magellanic Cloud-calibrated thinning track (the broken line), not unexpectedly for a halo PN with low metallicity. Further data are needed to investigate the reasons for a different thinning sequence, and future Gaia data releases may disclose some interesting developments in this direction.

\section{Discussion}

The Gaia parallaxes for Galactic CSPNe available in DR1 depict distances that are compatible, within the uncertainties, to spectroscopic distances in the literature. While there are still too few calibrators, and their uncertainties remain relatively high, the relation between physical radius and surface brightness seems already well defined (see Fig. 2) at all surface brightnesses of the parameter range. By adding to the calibration the PNe whose CS trigonometric parallax is available in the literature, we found ${\rm log}(R_{\rm pc})= -(0.454\pm0.043)\times {\rm log}(Sb_{\rm H\beta})-(7.274\pm0.610)$, which is  now applicable to a significantly wider range of the physical variable.

Comparably, the fit for the brightness temperature distance scale, ${\rm log}(R_{\rm pc})=-(0.429\pm0.041)\times{\rm log}(T_{\rm b})-(0.853\pm0.051)$, show a broad agreement with similar fits that exist in the literature, within the uncertainties; naturally, the TGAS parallaxes will anchor the relation at large $T_{\rm b}$ once the uncertainties will be lowered.

\begin{figure}
   \centering
  \includegraphics[width=\hsize]{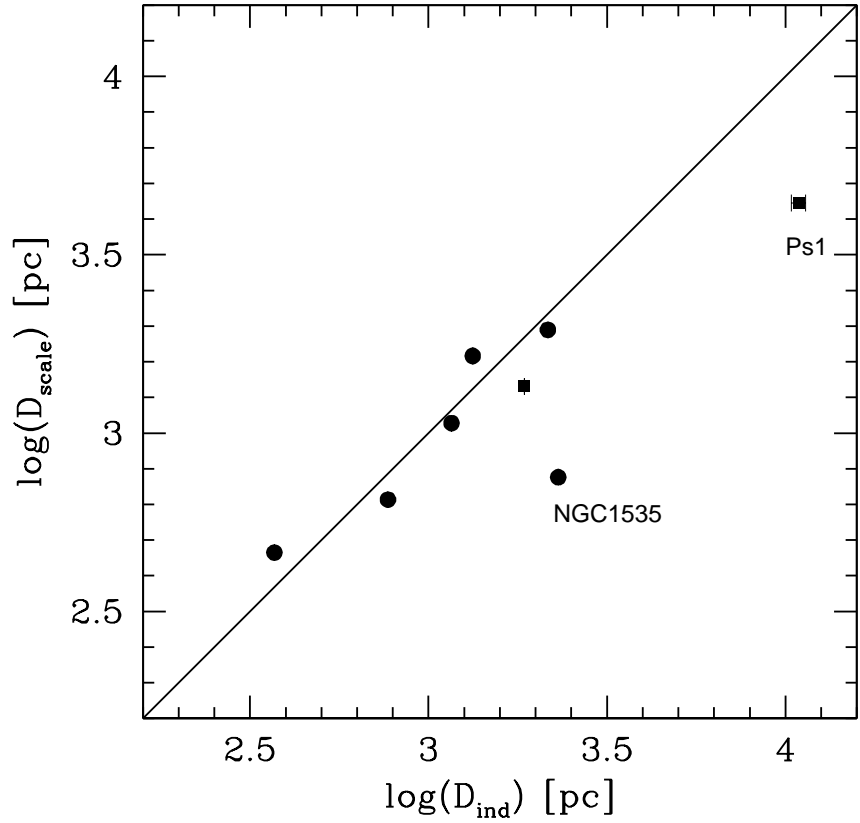}
\caption{Comparison between distances from the scale (Eq.~1) and reliable independent distances from the literature (see text,). independent distances, given in Table 5 for this comparison, are from cluster membership (squares) and spectroscopic parallax (circles).  All PNe with parallax distances used in the calibration of the scale have not been plotted here.  Discordant data points have been labeled with the PN name. The line is the 1:1 relation.}
\end{figure}

An initial assessment of the Gaia scale can be done by comparing distances derived from our preliminary scale (Eq. 1) with the independent distances. In Figure 5 we show the distances from our scale against independent distances determined by different methods. For this comparison we needed an external database which does not include any of the trigonometric parallax calibrators, either from TGAS or other parallaxes. We used the independent distances from two reliable methods: Spectroscopic parallaxes (Ciardullo et al. 1999);  and cluster membership distances from Chen et al. (2003, NGC 2818) and Otsuka et al. (2015, Ps~1). Parameters to derive the scale distances for these PNe are in Cahn et al.~(1992), and Stanghellini \& Haywood (2010). To help the reader, all of the data necessary to generate Fig. 5 are summarized in Table 5. We find that the scale gives results that compare reasonably well to the independent distances. The linear correlation coefficient between the scale and independent distances, in the log, is 0.88.

NGC~1535 (V$_{\rm CS}$=12.11, Ciardullo et al.~1999) and Ps~1 (V$_{\rm CS}$=14.73, Alves et al.~2005) are two evident exceptions to the statement above, as both their independent distances appear reliable. If cluster membership places  Ps~1 at $\sim$ 11 kpc, and therefore much further away than any of the sources used to derive Eq. 1, more intriguing is the case of NGC~1535, whose spectroscopic distance is much closer to the distances reached by the DR1 parallaxes utilized in the distance calibration. Notice that the 
Sb$_{\rm H\beta}$ fluxes of both sources fall well within the brightness range covered by our trigonometric calibrators.  Then, these two objects will be very special entries in the next Gaia data release for an immediate first  
re-assessment of the distance scale established here.

The forthcoming data releases from Gaia will provide parallaxes of  stars with V$<$15 with a precision of 0.03 mas (Lindegren et al.~2016). From Acker et al.~(1992) we found that there are $\sim$50 Galactic PNe whose CS magnitude V$_{\rm CS}<15$ and whose statistical distance is estimated (by means of the SSV scale) to be smaller than 3000 pc. For this group of PNe, Gaia will provide final parallaxes to better than 10$\%$ relative uncertainty; for another $\sim$40 PNe the estimated parallax relative uncertainty will be of the order of 20$\%$. 

The new Gaia data will be utilized to help building volume-limited samples on which to concentrate multi-wavelength surveys of PNe and their central stars. As recently discussed by Frew (2016), new astrometric and spectro-photometric volume-limited surveys are indeed needed to shed light on PNe evolutionary pathways.

The present work, although preliminary, sets the stage for confronting the expectations of the upcoming Gaia data in definitely constraining the PNe distance scales.

\section{Conclusions}

This paper presented an initial study of the Gaia TGAS parallaxes of CSs of Galactic PNe. We studied all PNe with available trigonometric distance from Gaia, collecting all the necessary physical parameters to infer their location on the two major distance scales, the {\it physical radius} and the {\it ionized mass} scales. 

By comparing the Gaia parallaxes with independent (and reliable) physical distances in the literature, we find good correlation within the uncertainties; however, the set is too limited for a comparative study and this must await for the next Gaia deliveries.
 
By building the PN {\it physical radius} distance scale with Gaia distances, and all reliable trigonometric parallaxes from the extant literature (whose physical parameters
have also been measured), we find a statistically tight linear correlation between the logarithmic physical radius and $H\beta$ surface brightness, with a linear correlation coefficient of -0.97 (see Fig. 2). A similarly high correlation is found between the physical radius and the brightness temperature of the PNe. The new sample studied here increases by $\sim$2 dex the domain of the parameter space with respect to older trigonometric parallax samples, making this set already very useful for distance scale analysis.

By studying the {\it ionized mass} distance scale for the Gaia sample (Fig. 4)  we realize that there are too few data constraining the scale for its two-branched shape, and that conclusions on this scale are premature. There are, however, indications that 
the thinning sequence, although 
similar to those of the older scales (e. g., Cahn et al.~1992), is shifted to higher inverse
surface brightness for the same ionized mass, suggestive of a different evolutionary path  of the PNe in the optically-thin sequence. However preliminary, this is the best
constrained thinning sequence so far, and it is very promising for future distance scale calibrations, especially when dealing with PNe in different environments.

We are looking forward to future data releases from Gaia to greatly increase the data sample, and lower the relative uncertainties of the parallaxes of Galactic PN central stars. Our expectation is that this new sample will lead to a much improved Gaia calibrated PNe statistical distance scale.

\section{Acknowledgements}

We acknowledge an anonymous Referee for carefully rewieving this paper. This work has made use of data from the European Space Agency (ESA)
mission {\it Gaia} (www.cosmos.esa.int/gaia), processed by
the {\it Gaia} Data Processing and Analysis Consortium (DPAC,
www.cosmos.esa.int/web/gaia/dpac/consortium). Funding
for the DPAC has been provided by national institutions, in particular
the institutions participating in the {\it Gaia} Multilateral Agreement.  We thank P. Marrese, with the ASDC at the Agenzia Spaziale Italiana (ASI), for providing us with the match of our PN list with the DR1 archive. L. S. acknowledges ALTEC (Torino, Italy) for support during the completion of this work.  Finally, B. B., M. G. L., and R. M. acknowledge the support of ASI through grant 2014-025-R.1.2015.


\begin{table*}
\caption{TGAS trigonometric parallaxes, and distances of the CSPNe}             
\label{table:1}      
\centering          
\begin{tabular}{ l l l c c  }    
\hline\hline       
  {PN~G}& {Name}& {ID}& {$p$} &log($D_{\rm p}$) \\
  &&&[mas]&[pc]\\

\hline
&&&&\\
038.2+12.0&     Cn~3-1&   4482121320558842880& 				1.932$\pm$0.656&    2.714$^{+0.180}_{-0.127}$  	\\
064.7+05.0& BD+303639&  2032744867997691008& 	0.279$\pm$0.432&    $\dots$\\
089.3--02.2&     M~1-77&  1971995510535755648& 					1.072$\pm$0.357&    2.970$^{+0.176}_{-0.125}$\\
104.8-06.7&M2~-54& 1990060349139814272&              -0.331$\pm$0.346&  $\dots$\\
165.5--15.2&     NGC~1514&  168937006671254144&                               2.286$\pm$0.239&   2.641$^{+0.048}_{-0.043}$\\
166.1+10.4&    IC~2149& 196996680850087936&         0.250$\pm$0.399&   $\dots$\\
215.6+03.6&   NGC~2346& 3109444653159703040&                                0.778$\pm$0.269&   3.109$^{+0.184}_{-0.129}$\\
272.1+12.3&   NGC~3132& 5420219727932911744&                                1.524$\pm$0.364&   2.817$^{+0.119}_{-0.093}$\\
311.0+02.4&     SuWt~2& 5870592987893097984& 					0.655$\pm$0.277&  3.183$^{+0.239}_{-0.153}$\\
315.1--13.0&   He~2-131& 5794858077215766656&      0.095$\pm$0.352&  $\dots$\\
316.1+08.4&   He~2-108& 5897352627007610240&     -0.007$\pm$0.872&  $\dots$ \\
321.0+03.9&   He~2-113&  5899715786733345536&    0.359$\pm$0.615&   $\dots$\\
331.1--05.7&      PC~11&  5899715786733345536&                                   0.741$\pm$0.682& 3.130$^{+1.100}_{-0.283}$\\
332.9--09.9&  He~3-1333&5917194997958036480&     0.190$\pm$0.652& $\dots$\\
334.8--07.4&  SaSt~2-12& 5923760662923825664&                                 0.310$\pm$0.243&  3.508$^{+0.662}_{-0.251}$\\
345.2--08.8&       Tc~1& 5954912369961887104&         0.219$\pm$0.489&  $\dots$\\
&&&&\\
\hline                    
  \hline                  
\end{tabular}
\end{table*}

\begin{table*}

\caption{Physical parameters of the PNe with TGAS parallaxes}
\label{table:2}
\centering
\begin{tabular}{ l c c c c}    
\hline\hline       
PN~G& $\theta$& 	log$(F_{\rm H\beta})$ &$c$ &$F_{\rm 5 GHz}$\\
&[arcsec]&  [erg cm$^{-2}$s$^{-1}$]& &[Jy]  \\ 

\hline
&&&&\\
Cn~3-1&  		2.25$\pm$0.12&      	     	-10.94$\pm$0.02&        	0.42$\pm$0.03 		 & 	0.067$\pm$0.007  \\
M~1-77&     	3.50$\pm$0.25 &      	     	-11.90$\pm$0.02$^a$ &      	1.34$\pm$0.02$^a$ 	&  	0.025$\pm$0.003    \\
NGC~1514&        66.0$\pm$8.0       &     	-10.98$\pm$ 0.03&        	0.96$\pm$0.40        & 	0.262$\pm$0.026$^a$ \\
NGC~2346&   	27.3$\pm$1.00      &	   	-11.26$\pm$0.02&       	 0.89$\pm$0.02$^a$          &  	0.086$\pm$0.016\\
NGC~3132&    	22.5$\pm$8.00        &	 -10.45$\pm$0.06&        	0.16$\pm$0.03           &	0.23$\pm$0.008 \\
SuWt~2&   	32.5$\pm$5.00      &           -12.35$\pm$0.02$^a$       & 0.64$\pm$0.02$^a$ & 0.007$\pm$0.001$^a$\\
PC~11&     	0.25$\pm$0.15      &		-11.48$\pm$ 0.02&        	0.89$\pm$0.02$^a$&                  0.011$\pm$0.001$^a$	\\
SaSt~2-12&  	0.44 $\pm$0.20    &	       	-11.19$\pm$0.02$^a$&		0.55$\pm$0.02$^a$  & 	0.082$\pm$0.008$^{a,b}$         \\

&&&&\\
\hline
\hline
\end{tabular}

$^a$ Assumed uncertainty, see text; 

$^b$ 5 GHz flux estimated from $H\beta$ intensity, see $\S$2.2.

\end{table*}

\begin{table*}
\caption{Independent distances of PNe with TGAS parallaxes}
\centering
\begin{tabular}{ l c l l}    
\hline\hline
Name& log($D_{\rm ind})$&  Method$^{a}$& Ref.\\
&[pc]&&\\
\hline
&&&\\

M~1-77&       3.398$^{+0.146}_{-0.222}$&    R&     HW88\\	
NGC~1514&  2.600&     SP&     MKU16, AM15\\
NGC~2346&    3.025$^{+0.121}_{-0.168}$&     R&   G86\\
NGC~3132&    2.886$^{+0.128}_{-0.182}$&      SP& C99\\	
SuWt~2$^{b}$&       3.362$^{+0.036}_{-0.040}$&         SP&  E10\\		
PC~11&              4.000$^{+0.025}_{-0.027}$&      SP&    P10\\
SaSt~2-12&        3.643&     SP&P04\\	

&&&\\
\hline
\hline
\end{tabular}

$^a$ R: Reddening; SP: spectroscopic parallax. 
References: AM15: Aller et al.~2015; C99: Ciardullo et al.~1999; E10: Exter et al.~2010; G86: Gathier et al.~1986; HW88: Huemer \& Weinberger 1988; MKU16: Mendez et al. 2016; P04: Pereira 2004; P10: Pereira et al.~2010.

$^b$ This distance is to the cold component of the binary system, thus not necessarily associated with the ionizing central star. We do not use this independent distance in the comparison of Fig. 1.

\end{table*}

\begin{table*}
\caption{CS trigonometric parallaxes in the literature, and physical parameters of their PNe}
\centering
\begin{tabular}{l l c c c c c}
\hline\hline
PN~G&Name& $p$&  $\theta$&log($F_{\rm H\beta})$ & $c$ & $F_{\rm 5 GHz}$ \\
& & [mas]& [arcsec]& [erg cm$^{-2}$ s$^{-1}$]& & [Jy]\\ 

\hline
&&&&&&\\
036.1--57.1&     NGC~7293&    4.56$\pm$0.49		&   	402.0&       -9.37$\pm$0.02 &       0.04 $\pm$0.02&		  1.292$\pm$0.026\\
060.8--03.6&     NGC~6853 &   2.64$\pm$0.33		&    170.0&     	-9.46$\pm$0.06    &    0.18$\pm$ 0.03	&	   1.325$\pm$0.019\\
063.1+13.9&     NGC~6720   &   1.42$\pm$0.55	&    34.6&       -10.08$\pm$0.03 &       0.29$\pm$0.04&		0.384$\pm$0.033\\
072.7--17.1&	    A~74         &  1.33$\pm$ 0.63	&    415.0  & 	-10.43$\pm$0.20$^a$& 	 $\dots$& 					$\dots$\\
158.5+00.7&  Sh~2-216 &	          7.76$\pm$0.33	&  $\dots$ & 	$\dots$& 			$\dots$&					 $\dots$\\
158.9+17.8 &      PuWe~1  &     2.74$\pm$0.31		&   600.0    &  	-10.85 $\pm$0.10 &      0.23$\pm$0.02&			0.085$\pm$0.017$^b$ \\
205.1+14.2& 	    A~21	    &  1.85$\pm$0.51 		&  307.5   &   	-10.40$\pm$0.20    &   	0.00$\pm$0.12&		    	0.327$\pm$0.065\\
215.5--30.8&          A~7       & 1.48$\pm$0.42  		&  382.0       & 	-10.11$\pm$0.20$^a$                   &  $\dots$&		    0.305$\pm$0.008\\
217.1+14.7&         A~24       & 1.92$\pm$0.34 		&  177.4    &  	-11.35$\pm$0.05  &      0.48$\pm$0.28&   		     0.036$\pm$0.004\\
219.1+31.2&         A~31       & 1.76$\pm$0.33  		&  486.0      &  	 -10.54$\pm$0.30     &   0.00$\pm$0.41&	   	0.102$\pm$0.020$^b$\\

&&&&&&\\
\hline
\hline
\end{tabular}
$^a$ Very uncertain flux, from original reference; assumed uncertainty, see text.

$^b$ 5 GHz flux estimated from $H\beta$ intensity, see text.

\end{table*}

\begin{table*}
\caption{Independent PN distances to be compared with those from our scale, and PN physical parameters.}
\centering
\begin{tabular}{l l l l l  r r r}
\hline\hline

PN~G&	
Name& 
log($D_{\rm ind}$)& 
Method& 
Ref$^a$& 
$\theta$& 
log($F_{\rm H\beta}$)& 
c\\

\hline
&&&&&&&\\

 238.0+34.8  &       A~33     &  3.064& SP& C99&    135     &    -11.3$\pm$0.05     &   0.38\\
  283.6+25.3   &    K1-22    &  3.124& SP& C99&    90.5   &     -11.42$\pm$0.10     &   0.12 \\
  329.3--02.8  &         Mz~2      &   3.334& SP& C99&  1.5   &     -11.65$\pm$0.02   &   0.5499\\
  206.4--40.5  &     NGC~1535   &  3.364& SP& C99&       9.2    &    -10.45$\pm$0.01  &       0.1\\
  261.9+08.5  &     NGC~2818   &  3.268& CM& C03&       20 &       -11.28&         0.3$\pm$0.04\\
  272.1+12.3 &      NGC~3132   &   2.886& SP& C99&    22.5  &      -10.45$\pm$0.06   &     0.16$\pm$0.03\\
  093.4+05.4 &      NGC~7008  &    2.568& SP& C99&      43  &      -10.86$\pm$0.05  &      0.84\\
  065.0--27.3  &         Ps~1     &    4.037& CM& O15& 1.8 &        -12.1$\pm$0.03     &    0.2$\pm$0.06\\

&&&&&&&\\
\hline
\hline
\end{tabular}
 $^a$ C99: Ciardullo et al.~1999; C03: Chen et al.~2003; O15: Otsuka et al.~2015.

\end{table*}

\end{document}